# Can unions impose costs on employers in education strikes? Evidence from pension disputes in UK universities

Nils Braakmann & Barbara Eberth[*]
Newcastle University

**Abstract**

The impact of strikes in educational institutions on employers remains understudied. This paper investigates the impact of education strikes in UK universities from 2018 to 2022, primarily due to pension disputes. Using data from the Guardian University Guide and the 2014 and 2021 Research Excellence Frameworks and leveraging difference-in-differences and regression discontinuity approaches, our findings suggest significant declines in several student related outcomes, such as student satisfaction, and a more mixed picture for student attainment and research performance. These results highlight the substantial, albeit indirect, cost unions can impose on university employers during strikes.

**Keywords:** University strikes, student outcomes

---

[*] Both Newcastle University, Business School, Newcastle upon Tyne, NE1 7RU, United Kingdom. Author contacts: nils.braakmann@ncl.ac.uk; Barbara.eberth@ncl.ac.uk

# 1. Introduction

The ability of unions and employers to impose costs onto each other is a central component in many bargaining models: Unions can call strikes, leading workers to withdraw their labour, while employers can lock out striking workers, leading to losses of wages and income. In many industries, strikes lead to a direct loss of output – if factory workers do not work on a specific day, that day's output is irrevocably lost. This is not true to the same extent in education settings: Education production is a gradual process with knowledge accumulating over time through multiple inputs – formal tuition, reading or discussions with peers (see, e.g., Hanushek, 2020, for an overview). Consequently, not being taught something on a given day does not mean that this knowledge can never be acquired. This, in turn, implies that a major lever usually used by unions during strikes might not work in the same way in education settings. So, are unions able to impose costs on employers in these settings?

First, for schools aimed at younger children, strikes disrupt childcare and thus the parents' ability to work, leading to (political) pressure to find an agreement. For example, in the only study of its kind, Jaume and Willén (2021) find large reductions in earnings, in particular for mothers, caused by teacher strikes in Argentina. In the context of secondary schools and – especially – universities this channel is, however, largely irrelevant as university students are adults and do not require constant supervision by their parents. Second, strikes cause a disruption to tuition and thus possibly a loss of learning. Additionally, this loss of learning can translate to lower qualifications being awarded, for example, if assessments are not adjusted for material that was not covered. Lower qualifications in turn can affect labour market prospects due to signalling – for example, when large employers demand certain minimum qualifications. There is indeed some evidence – again from schools – that teacher strikes can affect academic achievement: Belot and Webbink (2010) and Alvarado, Soler and González (2021) find effects of teacher strikes on the academic achievement of affected pupils in Belgium and Colombia respectively. Additionally, Jaume and Willén (2019) show that teacher strikes in Argentina led to large earning losses for affected pupils at ages 30 to 40, suggesting that there might be long-term losses of human capital – at least for the relatively long strikes observed in their data.

In this paper, we add to this small literature by presenting – to the best of our knowledge for the first time – evidence on the effects of education strikes at universities. Universities differ from schools in many important dimensions, which could lead to different effects of strikes. For example, a key component of many higher education courses is teaching the ability to (re-)acquire knowledge if it becomes necessary. For example, an HR advisor with an undergraduate degree in law would probably want to look up relevant current employment law when faced with a particular case – making it possibly irrelevant whether a lecture on this particular topic was cancelled due to industrial action a decade ago. In contrast, schools generally teach more basic and fundamental skills such as literacy. Furthermore, in the context of the strikes we consider, many universities adopted emergency regulations and awarded a "normal" distribution of marks and final degree classifications, which could attenuate concerns around signalling effects in the labour market. Nevertheless, it is possible that employers might consider these degree awards to be less reliable than degrees awarded in non-strike years – or at least that students fear that this might be the case. We indeed find evidence that strikes adversely affect "customer" opinion and outcomes, thus allowing unions to impose an, albeit indirect, cost on employers.

In addition – and unlike schools –, universities also conduct research, potentially giving unions another lever to use during strikes. However, incentives for individual staff to participate in a strike targeting research activities are potentially very different from incentives to participate in an education-focused strike. In many disciplines, research outputs are a major determinant of individual career progression and mobility as well as a very visible measure of individual productivity. If we think about this setting through a classic exit-voice framework (e.g., Hirschman, 1970; Freeman, 1980; Boroff and Lewin, 1997; Bender and Sloan, 1998), we can view strike activity as workers exercising their voice option. However, if workers also prepare for a possible exit – either of their individual employer or UK academia as a whole – in case their grievances are not resolved, it seems possible that striking workers would invest more effort on research activities to improve their individual CVs. In line with these arguments, we indeed find little evidence that universities affected by strikes saw changes in their performance in the Research Excellence Framework, a major assessment of research performance of UK universities conducted every 6-7 years (in our case in 2014 and 2021).

The setting we exploit are long-running strikes in UK higher education, triggered initially by changes to the Universities Superannuation Scheme (USS) – a large pension fund covering employees in the older UK universities[1] – and remaining unresolved during 2018 to 2022. During this period, the University and College Union (UCU) decided to ballot staff at individual universities on the possibility of industrial action, leading to within-year variation in strike action across different universities. Combining ballot results with data from a widely published national league table for undergraduate degrees for the academic years 2012/13 to 2021/22, the Guardian University Guide, we begin by exploiting the variation in strike action in a difference-in-difference setting with differential timing, i.e., we compare changes to the outcomes of universities subject to strike action to non-striking universities in the same year. Our approach controls for both individual heterogeneity at the university level – which is important as USS covers older and higher ranked universities, so that the bulk of strike activity is concentrated in those institutions – as well as temporal shocks common to all universities – such as the widespread move to online-learning during the COVID-19 pandemic. However, a concern are possibly localised effects of COVID-19, such as differences in COVID-restrictions between the constituent nations of the UK or the tier system of non-pharmaceutical interventions that was in operation in England from October 2020. We use the subset of universities located in municipalities[2] with multiple universities for regressions including local authority-by-year fixed effects, i.e., comparing striking and non-striking universities in the same area and year to gauge the size of an eventual bias caused by these locally-varying conditions.

---

[1] A common distinction in the UK is between pre- and post-1992 universities. The latter refer to former polytechnics that were granted university status through the Further and Higher Education Act 1992. They are often more teaching-focused and less research-intensive than the older universities. Due to their roots, they are also covered by a different pension scheme – the Teacher's Pension Scheme (TPS). However, staff who are members of USS can remain covered if they later join a university covered by TPS.

In a first step, we model the effects of strikes taking place during students' final year at university. In many UK universities the final year of an undergraduate degree carries a higher weight than earlier years when it comes to final marks.[3] Additionally, students fill in the National Student Survey in their final year, so that strikes in that year could have a large effect on measured satisfaction. In essence, in this setting we allow universities to switch into and out of treatment and model the short-term effect of the strikes on cohorts directly affected by them. We find somewhat mixed evidence for an immediate effect of the strikes: Depending on the estimator used, we find drops in overall league table positions as well as declines in student satisfaction. There is little evidence for a change in value-added – attainment measured by final degree classifications but adjusted for differences in intake qualifications.

However, an alternative way to think about the industrial action in the UK during this time is as one ongoing industrial dispute that remains unresolved and occasionally enters "hotter" periods with strikes. In this view, it seems possible that the first time a university enters strike action is pivotal – at this point, unionised staff are sufficiently motivated to enter industrial action and as the conflict is unresolved their goodwill and morale could erode over time. Given the fact that academics generally have a large amount of autonomy over their working lives, such drops could plausibly lead to changes in effort, leading to changes in teaching quality and the wider educational experience. Additionally, there might well be a cumulative effect of strikes on lost learning for cohorts affected by strikes during multiple years of their degree. To account for this possibility, we also model universities as treated in all years after they first experienced a strike. We indeed find evidence for larger effects in this setting – about twice as large in absolute value as for the first treatment definition: Universities drop on average 10 places in the Guardian ranking (from a pre-strike mean of 43), the composite score used by the Guardian drops by about 5 points (from a pre-mean of 66) and the proportion of students satisfied with teaching and the course as a whole drop by 1.6 and 1.9 percentage points respectively (from pre-means of 87% and 86%). Value added again remains unchanged. These findings are robust across a range of difference-in-difference estimators, including two-way fixed effects, estimators that are robust to heterogeneous and time-varying treatment effects such as Chaisemartin and d'Haultfœille (2020), Callaway and Sant'Anna (2021) and a synthetic difference-in-difference estimator recently proposed by Arkhangelsky et al. (2021) that provides additional robustness against common trend violations by combining elements of difference-in-differences with synthetic control methods (e.g., Abadie and Gardeazabal, 2002; Abadie et al., 2010).

---

[2] Municipalities in this context are local authority districts, the basic level of local government in the UK (roughly equivalent to US counties). Local authorities usually consist of a city or amalgamations of smaller towns and rural areas. London as a special case is split into 32 boroughs each designated as a Local Authority. There are a total of 348 Local Authorities in England and Wales, but not all contain one or multiple universities.

[3] For example, in a three-year degree a typical weighting scheme of module marks for a student's final degree classifications is 0 for modules in the first year, 1/3 for the second year and 2/3 for the final year.

In a next step, we estimate dynamic difference-in-differences, i.e., a series of event studies. We find that outcomes in universities entering strikes evolved similarly to those in other universities in the years prior to the strike, but experienced sharp and accelerating declines in outcomes in the three years following their first strike. We also conduct randomisation inference as a placebo test, where we randomise strike action across time and universities. These suggest our main estimates are well outside the effect range expected by chance. Next, we consider treatment effect heterogeneity across cohorts defined by the timing of the first strike. We find that the estimated effects are driven by those universities who experienced strike action from the very beginning of the conflict, i.e., from 2018, with later cohort experiencing much weaker and statistically insignificant drops.

We corroborate our main results by exploiting voting thresholds in a regression discontinuity design. Our setting is similar to DiNardo and Lee (2004) who exploited the fact that union recognition in the US changes discontinuously at 50% of votes in favour in a ballot. This allows for the comparison of firms where a union was just recognised with firms where recognition just failed. In our case, there are in principle two relevant thresholds: Industrial action requires a majority of cast votes in favour of strike action as well as a 50% turnout (Trade Union Act 2016). However, in our case the former threshold is practically irrelevant with the overwhelming majority of ballots reaching the required majority. In this case the union's ability to call a strike essentially hinges on reaching 50% turnout, allowing us to focus on this threshold. We implement our design using local polynomial regression discontinuity estimators with robust bias-corrected confidence intervals and inference procedures (Calonico, Cattaneo and Titiunik, 2014; Calonico, Cattaneo and Farrell, 2018; Calonico, Cattaneo, Farrell and Titiunik, 2019; Calonico, Cattaneo and Farrell, 2020). The resulting estimates identify similar parameters to our first difference-in-differences strategy, i.e., the effects of students being exposed to strike action in their final year. While not having large statistical power due to the number of observations near the threshold, we find estimates of identical signs and similar size to our main estimates.

Finally, we consider changes to research performance using data from the 2014 and 2021 Research Excellence Frameworks. Relying again on a difference-in-difference estimator, we find little evidence for changes in the research performance of affected universities. As argued earlier, however, we do not view this as necessarily good news for universities but suspect that this might be due to employees' focusing strike effort on teaching rather than research, possibly in preparation for an exit of their employer or UK academia.

The remainder of the paper is organised as follows: Section 2 presents a brief chronology of strikes in UK higher education and other relevant background. The data is introduced in section 3, followed by the empirical strategy in section 4. Results can be found in section 5, section 6 concludes.

## 2. Institutional background

*2.1 Workplace unionisation in UK higher education*

In general, if a union is recognised by an employer, it will represent all employees in specific groups, for example, all steelworkers or all administrative workers, regardless of whether these are union members. In our case, the University and College Union (UCU) represents academic employees at universities. Employers will not generally be aware whether a specific employee is a union member as there is no legal requirement to inform an employer about membership. In addition, union membership is a specifically protected characteristic under data protection legislation (under both the 1998 and 2018 Data Protection Acts as well as UK GDPR).

While not relevant in our case, as the unionisation status of universities does not change during the period of this paper, there are essentially two possibilities for a workplace to become unionised, voluntary and statutory recognition. In either case, the process begins with a written request for recognition by the respective union to an employer. At this stage, the employer can accept the request and enter collective bargaining with the union. If the employer rejects this request, the union can apply for statutory recognition to the Central Arbitration Committee (CAC), which essentially involves providing proof of sufficient membership and that a majority of employees supports union recognition.

In the case of UK higher education, UCU negotiates with two employer bodies: Universities UK (UUK) over matters related to the USS pension scheme and the Universities and Colleges Employers Association (UCEA) over matters related to pay and working conditions. In the case of industrial action due to national negotiations, UCU has a choice to either ballot members at each employer (university) individually or ballot members collectively at a national level. In either case, legal strike action is only possible if a majority of cast votes is in favour of it and if 50% of eligible members voted in the ballot (Trade Union Act 2016). During 2018/19 to 2021/22, i.e., the years of strike action that we focus on, UCU chose to ballot members at universities individually, which creates within-year variation in the universities that are affected by strike action. In 2022/23, strike action was decided by a national ballot instead.

*2.2 A brief history of strikes in UK higher education*

While our paper focuses on period of heave and prolonged industrial action from 2018 onwards, it is worthwhile to outline the history of the dispute and some earlier national ballots on strike action on pay in 2013/14 and an earlier pension dispute in 2014/15.

The 2013/14 pay dispute arose from a pay offer by UCEA to employees of only 1% in 2013/14. At this time employees including union members had seen a real pay cut by 13% since 2008. UCU balloted its members in September 2013 to take industrial action and industrial action short of a strike in response to the 1% pay offer. 62% of members voted in favour of strike action and 77% voted in favour of action short of a strike. The resulting strike action was relatively limited with two strike days in October and November 2013 and three 2-hour walk outs in January and February 2014. In February 2014 UCU announced a marking boycott to commence at the end of April if no better offer was forthcoming. In April 2014 UCEA offered a 2% pay increase for 2014/15, which was accepted by UCU members in early May.

Alongside the pay dispute, UUK proposed changes to USS in 2014 (e.g., Times Higher Education, 2014). The resulting conflict with UCU highlighted some of the issues that reappeared as part of the later pension disputes – disagreements about the valuation methodology for the scheme's assets and liabilities, the affordability of pension promises and what changes to pension entitlements were necessary. In October 2014, UCU balloted members in 67 universities for industrial action and to support a marking boycott which was to commence in early November if negotiations did not bring about satisfactory progress. The MAB was postponed after it was agreed that negotiations should take place during December and January to consider further proposals put forward by employers and alternative proposals by UCU. This resulted in a newly developed joint proposal for reform of the USS which UCU put to its members in a consultative ballot at the end of January 2015 (Times Higher Education, 2015).

The next national ballot took place in 2016 over pay. Employers offered a 1% uplift. UCU members again rejected the offer, resulting in two strike days at the end of May. The dispute was settled in November 2016 when UCU members accepted an uplift of 1.1%

The disputes that we focus on began in 2017/18 when the USS valuation estimated a deficit of £17.5 billon. In response, major changes to the scheme were proposed, including a closure of the defined benefit part of the scheme. Unlike the earlier disputes that were usually short, involved very limited strike action and were resolved within a few months, the resulting industrial action ended only in October 2023 after a total of 69 strike days with a reversal of the cuts to 2017 levels (e.g., Financial Times, 2023). Alongside the pension dispute, a dispute labelled the "Four Fights" by UCU commenced over pay and working conditions focussing in particular on pay levels, gender and minority ethnic pay gaps, staff workloads and insecure contracts. Unlike the earlier disputes, UCU also decided to ballot individual employers rather than conducting an aggregate/national ballot. The dispute also attracted high media attention as evidenced in Figure 1 which shows media mentions of higher education strikes from 2012 onwards.

(Figure 1 about here.)

These disputes led to an initial 14 days of strike action in 64 universities during February and March 2018, followed by a second wave of strike over the period 25th November and 4th December 2019 and a further 14 days staggered over the period of 20th February to 13th March 2020 on both the pensions and Four Fights dispute. The Covid 19 pandemic put a hold on industrial action given the lockdowns and resulting online teaching and remote learning for students but crucially, the worsening economic outlook had a detrimental impact on USS pension funds. The schemes valuation in March 2020 estimated a large and significant deficit of £14.1bn and resulted in further declines in pension benefits for members of the USS starting from April 2022 alongside increased employer contributions. UCU responded by further ballots for industrial action to restore benefits and a revaluation of the financial health of the pension scheme.

The start of the academic year 2021 saw a return to present in person teaching and learning and UCU balloted members again to resume industrial action over pension and the Four Fights. Strike days took place over the periods 1st – 3rd December 2021, 14th February – 2nd March 2022 and 21st March – 1st April 2022. For the first strike period, 58 Universities were on strike. Alongside this, members at 64 Universities started action short of a strike. After a re-ballot of those institutions that did not meet the threshold, UCU members at 68 Universities were on strike during February and March 2022.

As the industrial action mandate concluded on 3rd May, a further national ballot at 149 Universities was already underway in March 2022 to extend the strike mandate through the remainder of the year and to prepare for a marking boycott across institutions. This resulted in 10 further strike dates and a marking and assessment boycott starting on 23rd May. However, the appetite for a marking boycott was low; only 19 local branches commenced with the marking boycott.

After the end of our observation period, UCU announced a further ballot in July 2022 that was to be conducted as an aggregated/national ballot. Three strike dates were set for November 2022 and a further 18 days of strike action during February and March 2023. The pension dispute ultimately ended in October 2023 with a restoration of the pension benefits to 2017 levels. The pay dispute ultimately ended with an imposed settlement after UCU failed to reach 50% turnout in a summer ballot for continued strike action in 2024.

**3. Data**

We combine data from three sources: Official ballot results for the various industrial ballots by UCU, which we obtained from UCU's website, data on university-level teaching-related outcomes from the Guardian University Guide, a national league table produced by the newspaper the Guardian that draws on a range of administrative sources, such as the National Student Survey (NSS), the Universities and Colleges Admissions Service (UCAS) or the Higher Education Statistics Agency (HESA), and data from the UK Research Excellence Frameworks 2014 and 2021, which are major assessments of the research performance of every UK university that are conducted every 6-7 years.

Not all data used by the Guardian refers to the year of publication. For example, the latest available data, the 2024 guide, uses NSS data from 2022, entry standards from the academic year 2021/22, student-staff ratios from 2021/22 and career prospects (measured as the proportion of students either in graduate-level jobs or in further study) from 2019/20 and 2020/21. In addition, some data – such as expenditure per student – is aggregated over two years. We focus on outcomes, specifically NSS results and value-added, that we can clearly allocate to a specific academic year and that are available until 2021/22, after which UCU switched to a national ballot. Value-added is a measure of attainment, based on whether a student reaches a "good" degree outcome – defined as an upper second or first class honours degree – weighted by their ex-ante probability based on their prior qualifications to do so. It is collapsed into a score ranging from 1 (low value added) to 10. For the NSS results, we focus on overall satisfaction and satisfaction with the teaching on the course – which should be particularly sensitive to strike action. These are expressed as percentages, the proportion of students expressing satisfaction. We also use the overall Guardian rank and the composite score calculated by the Guardian (ranging from 0 to 100 with higher values indicating better outcomes). Both prospective students and staff in UK universities generally care about their ranking and ranking scores have been shown to influence undergraduate applications (Chevalier and Jia, 2016). Overall, we use data for 123 universities over 10 academic years (2012/13 to 2021/22). The panel is balanced for 115 universities with the majority of the eight remaining institutions entering the league table at some point after 2012/13.

To measure research-related outcomes, we use data from the 2014 and 2021 Research Excellence Frameworks. These are conducted every 6-7 years and are used to allocate research funding ("QR funding") to institutions. They involve peer reviews of (a) research outputs, such as journal articles or books, which are read by specialist panels for each discipline ("Unit of Assessment"), (b) an institution's research environment, based on metrics and a narrative statement provided by each institution, and (c) the real-world impact of an institution's research, based on written case studies provided by each institution. Scoring in each of the categories uses the same rating ranging from 0 (unranked) to 4 (world-leading). We focus on an institution's overall "GPA", calculated as the weighted average of these scores where the weights are the proportions of "elements" (outputs etc.) that achieved each score. We also compare research power, which is each institution's GPA multiplied by the number of full-time equivalent staff submitted and scaled so that the best outcome equals 100. Research power is usually interpreted as measuring an institution's scale as well as quality and can been viewed as approximating the size of the research funding block grant the institution can expect to receive. Specific rules, for example, around which members of staff needed to be submitted and how many outputs had to be submitted per member of staff, changed between 2014 and 2021. While we control for common time factors in our estimates, we also focus on institutions' relative positions by looking at their GPA and research power ranks relative to other universities, which should be more robust to specific rule changes.

Industrial action ballots are conducted by an independent organisation, Civita Election Services on behalf of UCU. Ballot results are published on the UCU website. In some years, such as 2021/22 some institutions were reballoted as an initial ballot either missed the necessary majority or the necessary turnout. In addition, during the 2019/20 and 2021/22 academic years there were separate ballots over USS and pay and conditions ("four fights"). We aggregate this data in the following way: We code an institution as being on strike if at least one ballot in an academic year reaches the necessary thresholds – a majority of votes in favour of strikes and a turnout of at least 50%.

For our RDD analysis we additionally need data for the percentage voting in favour of strikes and the turnout. In the case of multiple ballots for each institution and academic year, we prioritise ballots leading to strikes. The underpinning logic is that if an institution was balloted four times in an academic year, one of which resulted in a mandate for strike action, the institution ended up in a strike, i.e., the three failed ballots are ultimately not decisive. If more than one ballot resulted in a mandate for strike action, e.g., if an institution voted in favour of strikes over both pensions and pay and conditions, we retain the one with the higher turnout.

Figure 2 presents histograms across ballots for the percent voting in favour of strikes (Panel (a)) and the turnout (Panel (b)). We can clearly see that few ballots miss the necessary majority in favour of strike action with most of the ballot well above 50% in favour. However, the picture looks much more balanced for turnout with a good proportion of ballots distributed around 50% turnout. In Figure 3, we plot the incidence of strikes against both variables with the respective threshold marked by a dotted line. We can see that strike activity is largely determined by reaching the turnout threshold with a large discontinuity in the probability of strike action at 50%. In contrast, there is little evidence for the same effect at the 50% voting threshold (Panel (a)).

(Figures 2 and 3 about here.)

In Table 1, we present descriptive statistics for all universities as well as split by being affected by strike action after 2018. The top panel compares outcomes at baseline, i.e., prior to 2018. Universities later affected by are generally better ranked, have a higher guardian score, higher student satisfaction, a higher value added score and a better performance in the 2014 Research Excellence Framework.

(Table 1 about here.)

## 4. Empirical strategy

At a conceptual level, we are interested in estimating the causal effect of strike action at a university in a given time period on outcomes for that university. If strikes were randomly assigned to universities, we could straightforwardly compare outcomes for universities affected by strike action to outcomes for those unaffected. However, as we have just seen, strikes affect on average older and higher ranked universities, which would bias naïve comparisons. Instead, we rely on two quasi-experimental identification strategies: Most of our estimates are based on a difference-in-differences strategy that exploits the fact that different universities experience strikes in different academic years (or not at all), which allows us to control for both university-specific unobserved confounders and common time trends. For the education-related outcomes, we also present supplementary evidence using a regression discontinuity design (RDD) similar to DiNardo and Lee (2004) that exploits voting and turnout thresholds and essentially compares universities that just ended up in strike action due to a threshold being crossed to universities where the ballot just failed.

*4.1 Education outcomes: Difference-in-differences*

We exploit the fact that different universities experience strikes in different academic years and model this relationship in a difference-in-differences setup, i.e., we want to estimate

$$y_{it} = \alpha_i + \gamma_t + \tau * strike_{it} + \varepsilon_{it}, \qquad (1)$$

where *i* denotes universities and *t* academic years. $\alpha_i$ are university fixed effects, $\gamma_t$ an academic-year fixed effect and $strike_{it}$ is a dummy variable indicating strike activity. We define this in multiple ways as explained below. We cluster standard errors at the university level to adjust for heteroskedasticity and possible autocorrelation. For the former, clustering on the panel (cross-sectional) identifier is generally the appropriate level in a panel regression with more than two periods and fixed effects (Stock and Watson, 2009). In addition, difference-in-difference estimates with multiple time periods often suffer from autocorrelation with a commonly suggested remedy to cluster at the group level (Bertrand, Duflo and Mullainathan, 2004).

There are multiple econometric concerns that need considering in this general setup. The inclusion of university and time fixed effects controls for various confounders. The USS dispute was concentrated largely in the older, higher-ranked and more research-intensive universities. The inclusion of university fixed effects controls for these unobserved confounders as well as the baseline differences observed in Table 1. Similarly, the inclusion of time effects controls for common shocks that hit all universities, such as the onset of the COVID-19 pandemic, differences in the size of the 18-year-old population or inflationary pressures. A possible concern in relation to COVID-19 is that experiences of students could differ due to different local restrictions. To attenuate these concerns, we re-estimate (1) focusing on those universities in local authorities with multiple universities, which allows us to include local authority-by-year fixed effects that flexibly control for unobserved factors varying at the local authority-by-year level, for example, local COVID-restrictions such as nightlife or hospitality closures.

Additionally, the model in (1) is essentially a difference-in-differences setting with differential timing. A series of recent papers (de Chaisemartin and d'Haultfœille, 2020; Callaway and Sant'Anna, 2021; Goodman-Bacon, 2021; Borusyak, Jaravel and Spiess, 2023) have shown that estimating (1) by OLS/two-way fixed effects regressions (TWFE) can suffer from a range of econometric issues. To briefly sketch the underlying problems, it can be shown that $\tau$ in equation (1) is a weighted average of the treatment effects estimated by all possible 2*2 differences-in-differences that can be formed from the data (e.g., Goodman-Bacon, 2021). Some of these comparisons involve using already treated units as controls for later-treated units, which can lead to problems in the presence of time-varying and heterogeneous treatment effects, in extreme cases leading to an estimate for $\tau$ that is of opposite sign to all underpinning treatment effects. In addition, OLS places higher weights on units treated towards the middle of the sampling period. New estimators proposed by de Chaisemartin and d'Haultfœille (2020), Callaway and Sant'Anna (2021) and Borusyak, Jaravel and Spiess (2023) all deal with these problems in slightly different ways. Two of the estimators – Borusyak, Jaravel and Spiess (2023) and Callaway and Sant'Anna (2021) – assume an irreversible treatment, i.e., a treated unit remains treated for all later periods. The estimator by de Chaisemartin and d'Haultfœille (2020) can deal with units switching into and out of treatment by essentially comparing units switching from D=d to D=d' to units remaining at D=d, for example, comparing units where the treatment dummy switches from 0 to 1 to units remaining at 0.

In a first step, we model the effects of strikes taking place during students' final year at university. In this setting we allow universities to switch into and out of treatment and model the short-term effect of the strikes on cohorts directly affected by them. We present estimates based on a TWFE regression as well as the instantaneous treatment effect from the de Chaisemartin and d'Haultfœille (2020) estimator. However, an alternative way to think about the industrial action in the UK during this time is as one ongoing industrial dispute that remains unresolved and occasionally enters "hotter" periods with strikes. In this view, it seems possible that the first time a university enters strike action is pivotal – at this point, unionised staff are sufficiently motivated to enter industrial action and as the conflict is unresolved their goodwill and morale could erode over time. Given the fact that academics generally have a large amount of autonomy over their working lives, such drops could plausibly lead to changes in effort, leading to changes in teaching quality and the wider educational experience. Additionally, there might well be a cumulative effect of strikes and lost learning for cohorts affected by strikes during multiple years of their degree. To account for this possibility, we also model universities as treated in all years after they first experienced a strike. We again present two-way fixed effects regression results. In addition, we present estimates based on Callaway and Sant'Anna (2021).

Difference-in-differences requires two identifying assumptions: Limited anticipation of the treatment and (conditional) counterfactual common trends relative to the untreated group. The former essentially implies that we can pinpoint the exact timing of treatment and rules out behavioural adjustment of later-treated units in the pre-treatment period. In our case, universities were obviously aware of (a) staff being balloted for potential strike action before the results were known and strike action began and (b) the potential of further strikes after 2018 given that the initial industrial action did not lead to a resolution of the dispute. Indeed, many universities passed emergency regulations that allowed them to use higher degrees of discretion when awarding degrees. However, there was only limited potential for mitigation when it comes to striking lecturers cancelling lectures. First, universities generally do not know in advance which workers are union members and which union members are actually participating in a given strike. Second, many higher education modules require specialist knowledge, which makes it difficult to replace striking academics at short notice. Third, out of the 521 ballots conducted from January 2018 to January 2022, only 249 (48%) resulted in strike action, leading to further uncertainty whether a specific university would actually be treated. Against this background, it strikes us as reasonable to assume that anticipation effects should not play a major role for the results. However, as a corroborating strategy we rely on the regression discontinuity design described in the next section that ultimately uses only ballots that could have realistically resulted in either outcome.

Counterfactual common trends implies that outcomes for treated and untreated units would have evolved in the same way had the treatment not occurred. As this assumption involved counterfactuals – the outcome for the treated in a world where the treatment did not

happen – it is not directly testable. We use two commonly used indirect tests for the robustness of our results. We begin by randomising the timing of strikes and the universities affected by them to generate placebo estimates, which provides a sense of the natural variability in the outcomes and the "treatment effects" one would obtain if there was no actual treatment. We do this 250 times and compare our actual estimates with this placebo distribution in a randomisation inference test. We also present event study estimates that allow us to compare pre-existing trends between universities later affected by strike action and those that were not. These also allow us to check for an eventual bias caused by differential strike action prior to 2018. As an additional robustness check we estimate synthetic difference-in-differences using an estimator proposed by Arkhangelsky et al. (2021). This estimator combines features of difference-in-differences with the reweighting of control units used by synthetic control methods (e.g., Abadie and Gardeazabal, 2002; Abadie et al., 2010), which essentially ensures that treated and control units followed common trends in the pre-treatment period. The estimator requires a non-reversible treatment, so that these estimates are comparable to the TWFE and Callaway and Sant'Anna (2021) estimates that consider universities to be treated after the first industrial action. It also requires a balanced panel, so estimates are based on those 115 institutions for which data is available for every year. Finally, we explore heterogeneity of estimates across cohorts of universities defined by the timing of the first strike and across years.

*4.2 Education outcomes: Regression discontinuity design*

A possible concern with the difference-in-difference strategy is that at least some universities might be able to predict whether they would be affected by strike action, for example, if they know that they have a very active or inactive UCU branch. To address these concerns, we use a supplementary identification strategy and focus on close ballots in a regression discontinuity design, comparing universities where a ballot just reached the necessary majority and turnout for a strike with those where a ballot just failed. In these cases, it seems credible that neither the union nor the university would be able to predict whether a strike would occur.

Strike action requires that a majority of union members vote in favour and at least 50% of eligible members cast a vote. At first glance, this setting looks like an RDD with two thresholds, however, in practice only the turnout threshold is relevant: As shown earlier in figures 1 and 2, the overwhelming majority of ballots reaches the required majority. In this case the union's ability to call a strike essentially hinges on reaching 50% turnout, which means that we can focus on this threshold. We implement our design using local polynomial regression discontinuity estimators with robust bias-corrected confidence intervals and inference procedures (Calonico, Cattaneo and Titiunik, 2014; Calonico, Cattaneo and Farrell, 2018; Calonico, Cattaneo, Farrell and Titiunik, 2019; Calonico, Cattaneo and Farrell, 2020).

Our design compares universities where a strike ballot just passed due the 50% being reached with universities where the ballot just led to a rejection of strike action. As the RDD uses only university-year-combinations where a ballot took place our initial sample is much smaller at 287 university-year combinations. However, the effective number of observations under optimal bandwidth selection is 25 to the left of the threshold and between 48 and 64 to the right. Technically, the design is fuzzy as a few universities reached the 50% threshold but had a majority of majority of union members voting against action. Given the small sample sizes, however, we choose to treat the design as sharp to improve efficiency – essentially ignoring these universities. As a consequence, the estimated treatment effects are likely too small as we do not scale the difference in outcomes at the threshold by the difference in the proportion of universities on strike.

*4.3 Research outcomes: Two-period differences-in-differences*

Our estimation strategy for research-related outcomes is largely similar to the difference-in-differences strategy described in Section 4.1 with one major change: As we only have two data points for each university from the 2014 and 2021 Research Excellence Frameworks, our setting is essentially a two-period difference-in-differences. In addition, only the 2018 and 2019 are relevant in this setting as the 2021 Research Excellence Framework predates the 2021/22 strikes. Similar to earlier, we estimate

$$y_{it} = \alpha_i + \gamma_t + \tau * strike_{it} + \varepsilon_{it}, \qquad (2)$$

where $\alpha_i$ is again a university fixed effect, $\gamma_t$ is a dummy variable for 2021 and $strike_{it}$ marks universities affected by strike action in either 2018 or 2019. We also provide separate estimates for universities first going on strike in 2018 or 2019. Given the shorter time period, we cannot estimate event studies and essentially have to assume that the common trend assumption holds.

## 5. Results

*5.1 Main education results and robustness*

Table 2 summarises our main results: The top panel (a) presents results for the immediate impact of the strikes, the bottom panel (b) presents results for the models where we consider universities to be treated after their first strike action. In both cases, the estimates obtained from two-way fixed effects regressions and the specialised estimators by de Chaisemartin and d'Haultfœille (2020) and Callaway and Sant'Anna (2021) are qualitatively and at least for panel (b) also quantitatively identical.[4]

(Table 2 about here.)

Panel (a) suggests drops in Guardian league table positions by between 2 and 6 ranks and corresponding drops in the overall Guardian score by between 2 and 3 points. Estimates on student satisfaction is more mixed with the TWFE estimates suggesting a drop by around 1 percentage point for teaching and course satisfaction and the de Chaisemartin and d'Haultfœille (2020) estimates suggesting smaller and statistically insignificant changes. Neither estimator suggests a large effect on value added. Panel (b) suggests much larger effects – generally around twice as large as those found in panel (a): League table positions worsen by 9 to 10 ranks, the overall score by around 4 to 5 points and NSS results by around 2 percentage points. These estimates are remarkably similarly across the TWFE, Callaway and Sant'Anna (2021) and synthetic difference-in-differences (Arkhangelsky et al., 2021) estimators.

How large are these effects in absolute terms? A useful benchmark is the within-variation in the affected universities in the pre-strike years, i.e., how much do university outcomes usually change even if there is no strike action. In our case, the within standard deviation of the Guardian rank from 2012-2017 is 8, 3 for the Guardian Score, 2 for the NSS results for satisfaction with the course, 1.5 for satisfaction with teaching and 0.5 for value-added. Overall, this suggests that effect sizes are approximately equal to between one and two standard deviations, which is sizeable.

As stated earlier there are two possible explanations for the larger longer-term effects: First, it could be that later student cohorts are more strongly affected due to the cumulative impact of strike action over multiple years (or the cumulative impact of strike action and COVID-19 related disruption). Second, it could also be the case that ongoing industrial action over the same issue without resolution led to a decline in staff goodwill leading to worse outcomes even in periods when no strike action takes place. The lack of an effect on value added could be due to mitigation measures by universities, such as changing assessments to only cover material that was actually taught or the use of partial marks for degree classifications.

(Figure 4 about here.)

---

[4] This pattern mimics a recent replication exercise of 38 papers published in the *American Political Science Review*, *American Journal of Political Science*, and *Journal of Politics* that found that estimators robust to differential timing and heterogeneous treatment effects generally did not reach different substantial conclusions than estimates based on two-way fixed effects regressions (Chiu et al., 2023).

This general pattern from Table 2 is confirmed in Figure 4 where we present event-study estimates. Universities experiencing strike action are generally similar to those that do not in the 5 years prior to the first strike action after 2018 – this also suggests that the earlier limited strikes do not bias our estimates. Subsequently, they experience a sharp and usually accelerating worsening of the respective outcomes, again with the exception of value added. Two years after the onset of the strikes, the estimates suggest a drop in the Guardian ranking by around 11 to 12 ranks, a decline in the overall score by around 5 and drops in the NSS scores by between 2 and 2.5 percentage points.

(Table 3 about here.)

In Table 3, we present results from the placebo tests. These basically confirm that any effects that were statistically significant in the main estimates are unlikely to have arisen due to chance.

(Figure 5 about here.)

We now explore heterogeneity across cohorts of universities where cohorts are defined by the timing of the first strike action. Results can be found in Figure 5. Across the four outcomes where we have observed negative effects previously, Figure 5 suggests that these negative effects are driven by universities that were affected by strike action from the very beginning of the pension disputes in 2018. In contrast, effects for universities that were first affected during the 2019/20 or the 2021/22 academic year are much smaller and statistically insignificant.

*5.2 Assessing biases due to COVID-19: Within-city comparisons*

A possible complication is that the period of strikes that we are considering is interrupted by the COVID-19 pandemic and the associated non-pharmaceutical interventions. These can have an independent effect on student satisfaction and experience, for example, because aspects of student life were affected by closures of hospitality and nightlife establishments. The restrictions differed between England, Northern Ireland, Scotland and Wales and, at least for a time, also between areas in England where they were based on local COVID infection rates. These local differences could bias our estimates if universities affected by strikes were also based in areas more strongly affected by COVID restrictions.

To test for this possibility, we exploit the fact that 56 universities are situated in local authorities with more than one university. This allows us to add local authority-by-year effects – non-parametric time trends that vary by local authority – and essentially compare striking and non-striking universities in the same city and year.

(Table 4 about here.)

Results can be found in Table 4. While standard error are usually larger than for the corresponding full-sample estimates in Table 2 – which is not surprising given the smaller sample size – point estimates are of a similar size and statistical significance is often unchanged. This suggest that differences in COVID-19 restrictions are not a major driver of the results.

*5.3 Regression-discontinuity design estimates*

Table 5 presents the regression discontinuity estimates based on 287 university-year combinations where industrial ballots were held, 151 of which resulted in strike action. The optimal bandwidth selection generally uses 25 observations to the left of the turnout threshold and between 48 and 64 to the right. These estimates link strike action in a given year to outcomes in the same year. They are thus comparable to the estimates from Table 2, panel (a). While effects are not very precisely estimated due to the small number of observations, point estimates generally lie between the TWFE estimates in panel (a1) of Table 2 and the de Chaisemartin and d'Haultfœille (2020) estimates in Panel (a2). We would interpret this as supportive evidence for our main estimation strategy.

(Table 5 about here.)

*5.4 Effects on research performance*

Table 6 presents our estimates for effects on research performance. Panel (a) compares all universities affected by strike action in either 2018 or 2019 to those which were not on strike, panel (b) splits the treatment into those which entered strikes for the first time in 2018 and those which first experienced strike action in 2019 and compares either to universities not on strike. Overall and across the four outcomes, there is little evidence for a substantial change to research performance. Points estimates are on balance suggestive of a drop in performance, however, these are small and statistically insignificant

(Table 6 about here.)

## 6. Conclusion

We investigated to what extent strikes in higher education affected university-level outcomes, such as student satisfaction and league table positions. We explored this question in the context of long-running pension disputes in UK higher education. Overall, our estimates suggest that the standing of universities in a major league table as well as the satisfaction of their students declined due to the strike. Our estimates also suggest that these negative effects are only partially driven by the direct disruption experienced by some students and instead persist and even accelerate over periods that are part of the same dispute but that do not experience direct strike action. We also find that the negative effects are concentrated in those universities that have been affected by strike action since the first year of the dispute in 2018. Possible explanations include the cumulative effects of disruption experienced by students across multiple years of their studies, the cumulative disruption due to the strikes and the COVID-19 pandemic and worsening relations between academic staff and their institutions leading to a loss of goodwill. We do not find much evidence for a decline in research performance. As argued earlier, however, we do not view this as necessarily good news for universities, but suspect that this might be due to employees' focusing strike effort on teaching rather than research, possibly in preparation for an exit of their employer or UK academia.

More broadly, our estimates suggest that unions are able to impose costs on employers in a sector where strike action does not lead to a direct and obvious loss of output. In fact, the unique nature of higher education, with its focus on fostering independent knowledge acquisition and its flexibility in terms of curriculum delivery, would have suggested an *a priori* reduced vulnerability to the effects of strikes. However, our estimates clearly suggest that even in this setting protracted labour disputes carry negative consequences, even thought they might be less directly visible.

**Table 1: Descriptive statistics**

| Variable | All universities | Universities without strike after 2018 | Universities affected by strikes after 2018 |
|---|---|---|---|
| *Strike and education-related data* | | | |
| *Pre-2018* | | | |
| Guardian rank | 60 | 84 | 43 |
| | (34) | (25) | (29) |
| Guardian score | 60 | 50 | 66 |
| | (14) | (9) | (12) |
| Satisfied with teaching (NSS) | 86 | 85 | 87 |
| | (3) | (3) | (3) |
| Satisfied with course (NSS) | 85 | 83 | 86 |
| | (4) | (4) | (4) |
| Value added score | 5 | 5 | 6 |
| | (1) | (1) | (1) |
| *Post-2018* | | | |
| Turnout (%) | 48.4 | 36.7 | 53.8 |
| | (11.9) | (8.9) | (8.9) |
| Proportion yes vote (%) | 75.3 | 68.6 | 78.3 |
| | (10.5) | (10.3) | (9.0) |
| Guardian rank | 60 | 79 | 48 |
| | (35) | (30) | (32) |
| Guardian score | 62 | 56 | 66 |
| | (12) | (10) | (12) |
| Satisfied with teaching (NSS) | 84 | 84 | 84 |
| | (3) | (3) | (4) |
| Satisfied with course (NSS) | 82 | 82 | 83 |
| | (5) | (4) | (5) |
| Value added score | 6 | 5 | 9 |
| | (1) | (1) | (1) |
| Observations | 1188 | 474 | 714 |
| *Research Excellence Framework data* | | | |
| *2014* | | | |
| Grade Point Average | 2.73 | 2.46 | 2.99 |
| | (0.39) | (0.34) | (0.34) |
| GPA rank | 64 | 89 | 39 |
| | (36) | (27) | (25) |
| Research Power | 155 | 54 | 251 |
| | (190) | (54) | (221) |
| Research Power rank | 62 | 87 | 38 |
| | (36) | (27) | (28) |
| *2021* | | | |
| Grade Point Average | 2.96 | 2.70 | 3.22 |
| | (0.39) | (0.35) | (0.23) |
| GPA rank | 64 | 90 | 39 |
| | (36) | (26) | (26) |
| Research Power | 166 | 68 | 262 |
| | (189) | (60) | (221) |
| Research Power rank | 63 | 86 | 40 |
| | (37) | (29) | (29) |
| Observations | 240 | 118 | 122 |

**Table 2: Difference-in-difference estimates**

|  | Guardian rank | Guardian score | Satisfaction with teaching (NSS) | Satisfaction with course (NSS) | Value added score |
|---|---|---|---|---|---|
| *Panel (a): Short term effects: Strike in respective year* | | | | | |
| *Panel (a1): Two-way fixed effects* | | | | | |
| Strike (1 = yes) | 5.73** | -3.10*** | -0.70** | -1.06*** | 0.01 |
|  | (2.40) | (0.77) | (0.29) | (0.39) | (0.11) |
| *Panel (a2): de Chaisemartin and d'Hautlfœille (2020)* | | | | | |
| Instantaneous treatment effect | 1.75 | -1.77** | 0.08 | -0.24 | 0.06 |
|  | (1.71) | (0.55) | (0.20) | (0.23) | (0.09) |
| *Panel (b): Longer term effects: After first strike* | | | | | |
| *Panel (b1): Two-way fixed effects* | | | | | |
| After first Strike (1 = yes) | 10.07*** | -5.46*** | -1.58*** | -1.87*** | -0.09 |
|  | (3.08) | (1.00) | (0.37) | (0.47) | (0.14) |
| *Panel (b2): Callaway and Sant'Anna (2021)* | | | | | |
| Overall ATT | 8.99*** | -4.42*** | -1.37*** | -1.82*** | -0.01 |
|  | (3.15) | (1.04) | (0.36) | (0.45) | (0.15) |
| *Panel (b3): Synthetic difference-in-differences (Arkhangelsky et al., 2021)* | | | | | |
| Overall ATT | 8.337*** | -5.354*** | -2.027*** | -1.702*** | -0.082 |
|  | (3.170) | (1.050) | (0.578) | (0.479) | (0.162) |
| Observations (b1 and b2) | 1,187 | 1,187 | 1,187 | 1,187 | 1,187 |
| Observations (b3) | 1,150 | 1,150 | 1,150 | 1,150 | 1,150 |
| University FE | Yes | Yes | Yes | Yes | Yes |
| Year FE | Yes | Yes | Yes | Yes | Yes |

*Notes:* Panels (a1) and (b1) display the regression coefficients from a TWFE regression, panels (a2) and (b2) the aggregated overall average treatment effect on the treated. */**/*** denote statistical significance at the 10%, 5% and 1% levels. Standard errors adjusted for clustering at the university level in parentheses. Data from 2012 to 2021, strikes beginning in 2018.

**Table 3: Placebo tests, randomisation inference, 250 replications**

| | Guardian rank | Guardian score | Satisfaction with teaching (NSS) | Satisfaction with course (NSS) | Value added score |
|---|---|---|---|---|---|
| *Panel (a): Short term effects: Strike in respective year* | | | | | |
| *Panel (a1): Two-way fixed effects* | | | | | |
| Strike (1 = yes) | c = 0 | c = 0 | c = 0 | c = 0 | c = 234 |
| | p=0.000 | p=0.000 | p=0.000 | p=0.000 | p=0.936 |
| *Panel (b): Longer term effects: After first strike* | | | | | |
| *Panel (b1): Two-way fixed effects* | | | | | |
| After first | c = 0 | c = 0 | c = 0 | c = 0 | c = 24 |
| Strike (1 = yes) | p=0.000 | p=0.000 | p=0.000 | p=0.000 | p=0.096 |
| Observations | 1,187 | 1,187 | 1,187 | 1,187 | 1,187 |
| University FE | Yes | Yes | Yes | Yes | Yes |
| Year FE | Yes | Yes | Yes | Yes | Yes |

*Notes:* Each cell displays the number of placebo estimates larger in absolute value than the observed effect, $c = |\tau| \geq |\tau^{Obs}|$ as well as the corresponding empirical p-value, $p = c/250$.

**Table 4: Within-city comparisons**

| | Guardian rank | Guardian score | Satisfaction with teaching (NSS) | Satisfaction with course (NSS) | Value added score |
|---|---|---|---|---|---|
| *Panel (a): Short term effects: Strike in respective year, two-way fixed effects* | | | | | |
| Strike (1 = yes) | 5.98 | -3.18*** | -0.60** | -0.94*** | 0.10 |
| | (3.94) | (1.29) | (0.33) | (0.47) | (0.14) |
| *Panel (b): Longer term effects: After first strike, two-way fixed effects* | | | | | |
| After first Strike (1 = yes) | 6.88 | -5.75*** | -1.51*** | -1.86*** | 0.04 |
| | (4.36) | (1.52) | (0.57) | (0.78) | (0.24) |
| Observations | 586 | 586 | 586 | 586 | 586 |
| University FE | Yes | Yes | Yes | Yes | Yes |
| Year-by-local authority FE | Yes | Yes | Yes | Yes | Yes |

*Notes:* Panels (a) and (b) display the regressions coefficient from a TWFE regression. */**/*** denote statistical significance at the 10%, 5% and 1% levels. Standard errors adjusted for clustering at the university level in parentheses. Data from 2012 to 2021, strikes beginning in 2018.

**Table 5: Regression discontinuity estimates using turnout threshold**

|  | Guardian rank | Guardian score | Satisfaction with teaching (NSS) | Satisfaction with course (NSS) | Value added score |
|---|---|---|---|---|---|
| ATT | 0.22 | -1.90 | -1.01 | -0.23 | -1.40*** |
|  | (14.59) | (5.12) | (1.62) | (2.28) | (0.49) |
| Observations | 287 | 287 | 287 | 287 | 287 |
| Effective Obs [left, right of threshold] | [25,56] | [25,64] | [25,50] | [25,48] | [25,50] |

*Notes:* Point estimates, standard errors in parentheses. */**/*** denote statistical significance on the 10%, 5% and 1% level respectively. Estimates based on local polynomial regression discontinuity estimators with robust bias-corrected confidence intervals and inference procedures (Calonico, Cattaneo and Titiunik, 2014; Calonico, Cattaneo and Farrell, 2018; Calonico, Cattaneo, Farrell and Titiunik, 2019; Calonico, Cattaneo and Farrell, 2020). Triangular kernel.

**Table 6: Effect on Research Excellence Framework outcomes, Difference-in-differences estimates**

|  | GPA | GPA rank | Research power | Research Power Rank |
|---|---|---|---|---|
| *Panel (a): Any strike in 2018 or 2019* | | | | |
| Affected by strike | -0.028 | 0.945 | -6.170 | 3.959** |
|  | (0.030) | (2.411) | (5.686) | (1.679) |
| *Panel (b): Effects by year of first strike* | | | | |
| First strike in 2018 | -0.021 | -0.037 | -8.154 | 4.761*** |
|  | (0.029) | (2.388) | (6.195) | (1.687) |
| First strike in 2019 | -0.062 | 5.952 | 3.952 | -0.128 |
|  | (0.061) | (5.194) | (7.159) | (3.420) |
| Observations | 240 | 240 | 240 | 240 |
| University FE | Yes | Yes | Yes | Yes |
| Year FE | Yes | Yes | Yes | Yes |

*Notes:* Panels (a) and (b) display the regressions coefficient from a TWFE regression. */**/*** denote statistical significance at the 10%, 5% and 1% levels. Standard errors adjusted for clustering at the university level in parentheses. Data from 2014 and 2021 REF exercises.

**Figure 1: Media mentions of university strike**

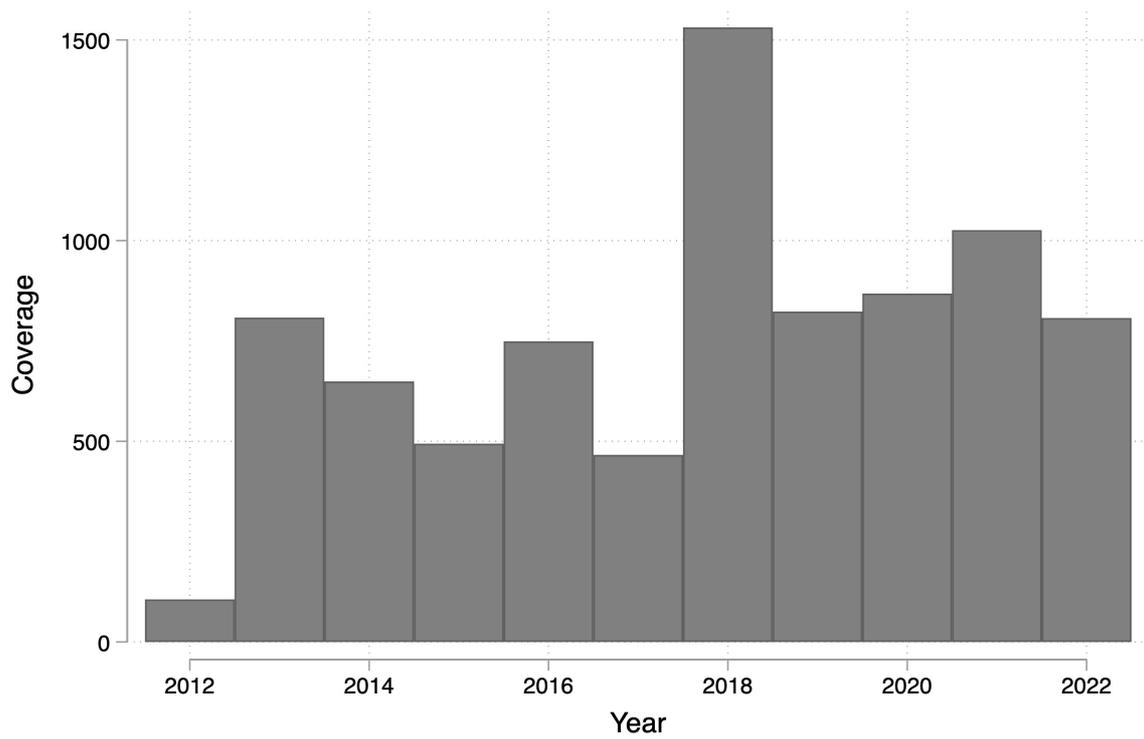

Note: Mentions of "university strikes" in news outlets in the UK from 09/2012 to 09/2022 from LexisNexis.

**Figure 2: Distribution of union ballot results**
Panel (a): Proportion of yes votes

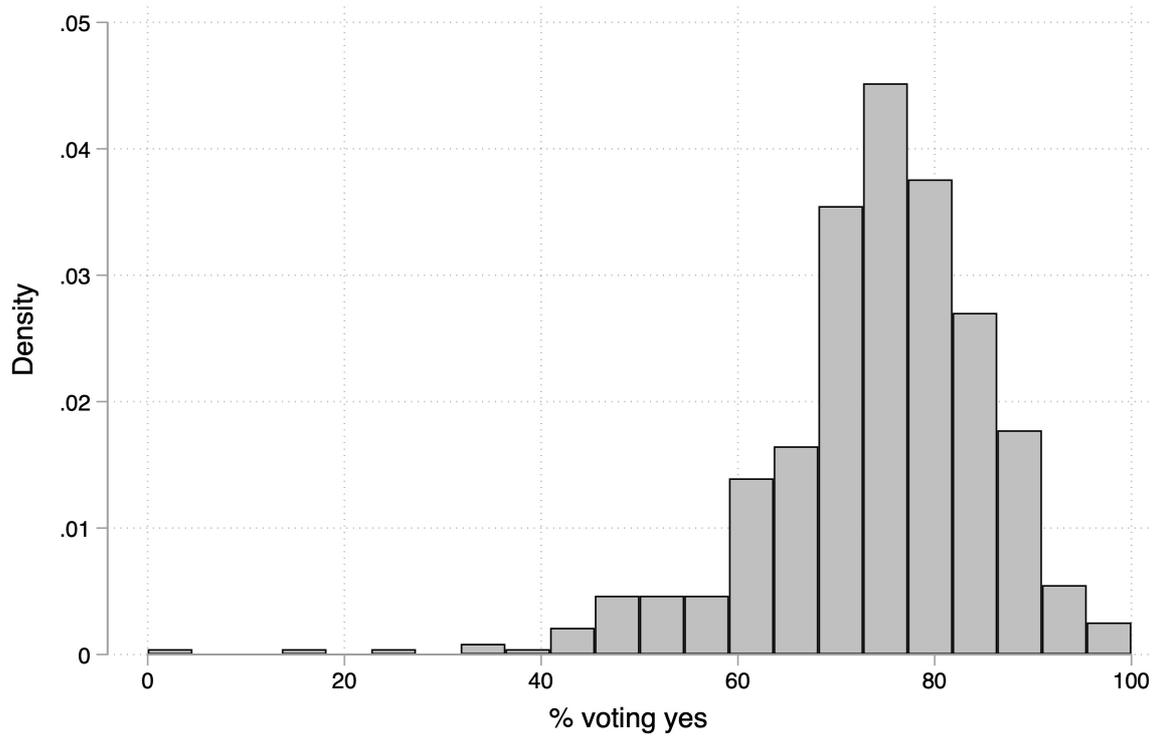

Panel (b): Turnout

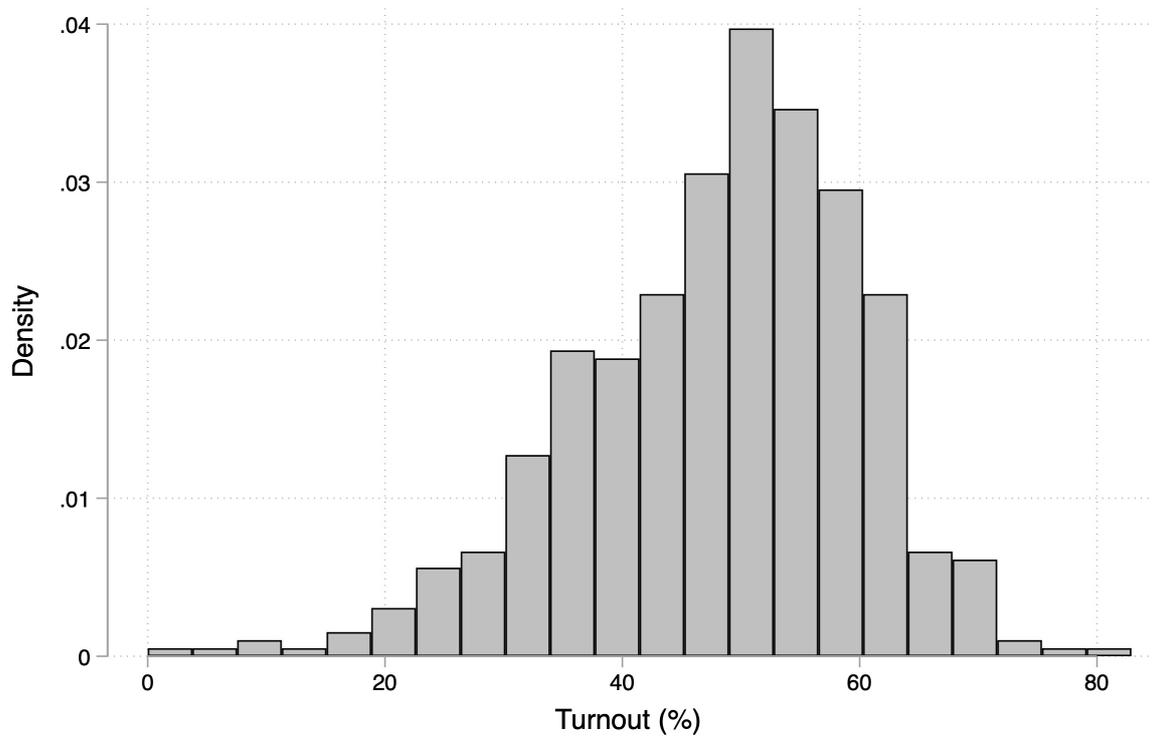

**Figure 3: Binned scatter plots: Strikes, yes votes and turnout**
Panel (a): Strikes and proportion of yes votes

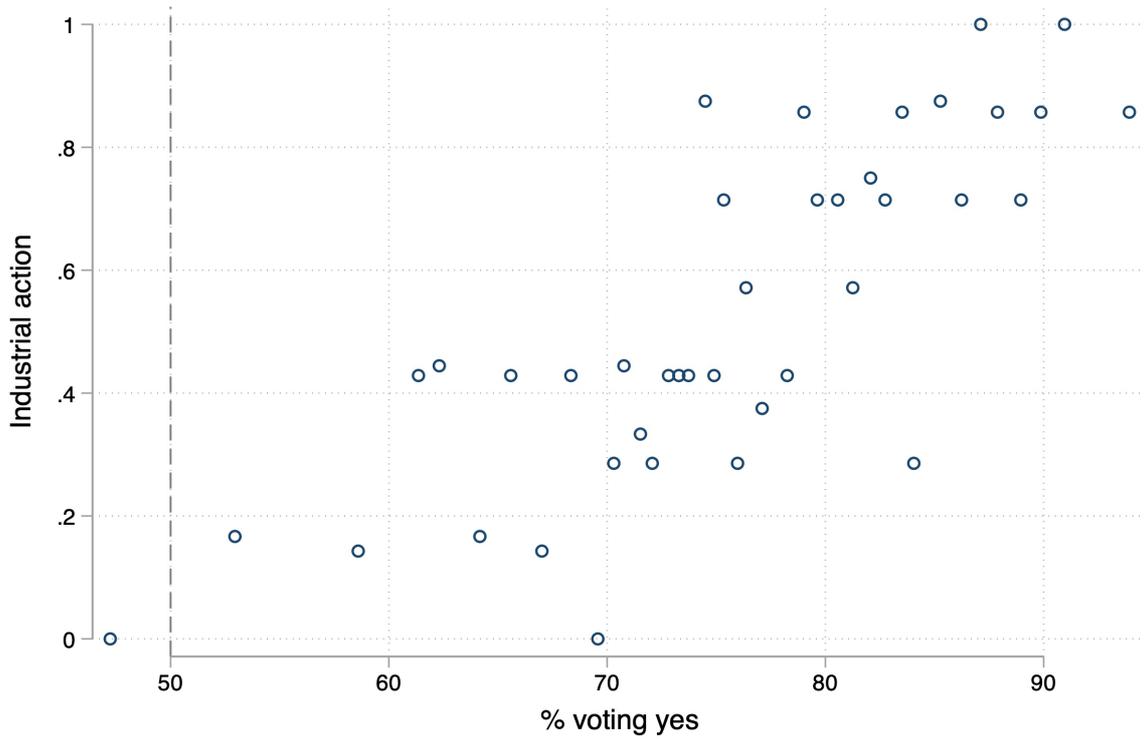

Panel (b): Strikes and turnout

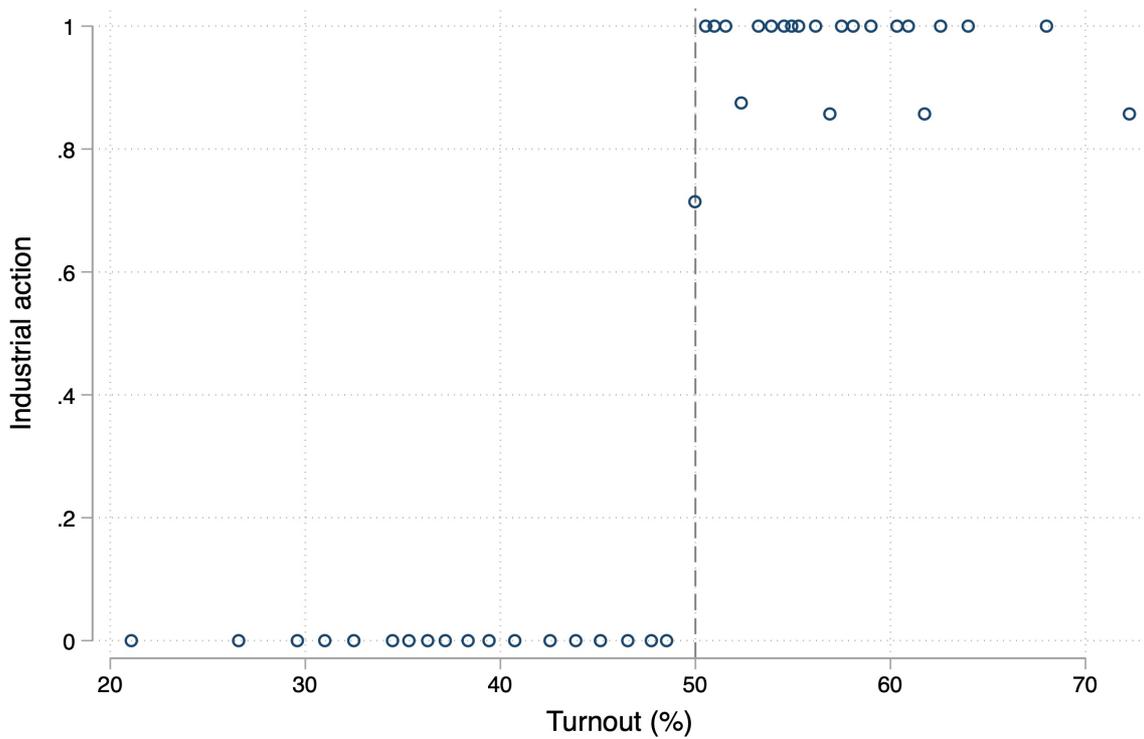

**Figure 4: Event study plots, Callaway and Sant' Anna (2021)**
Panel (a): Guardian Rank

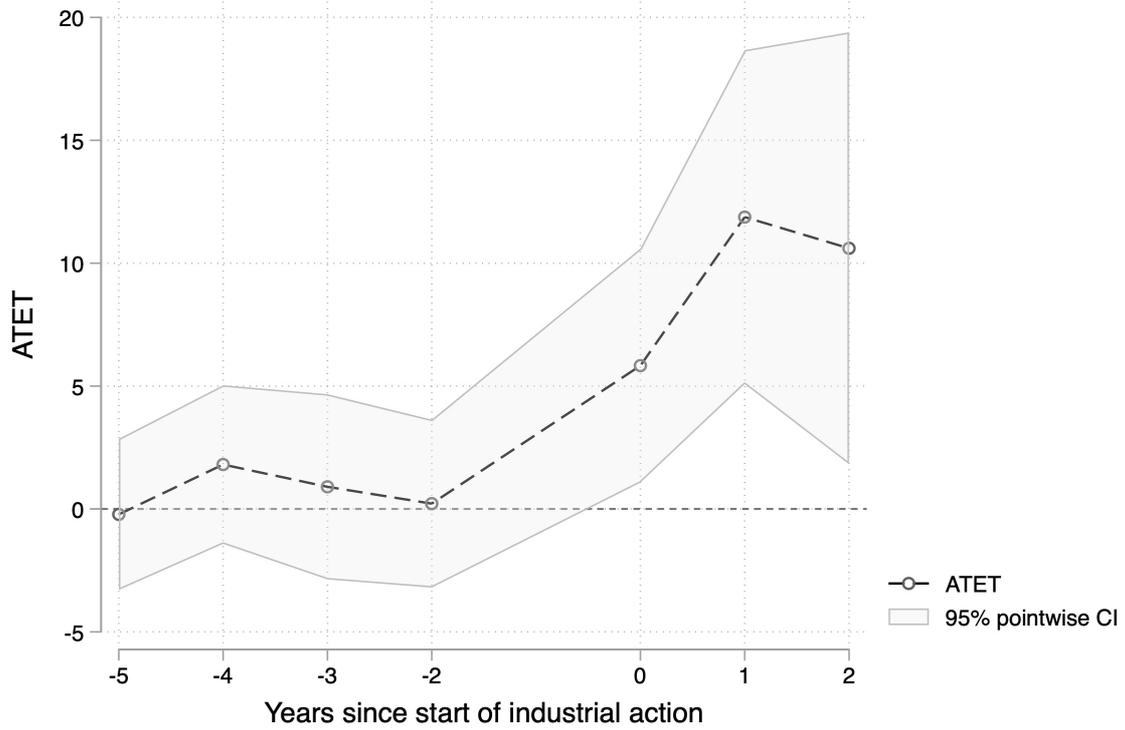

Panel (b) Guardian Score

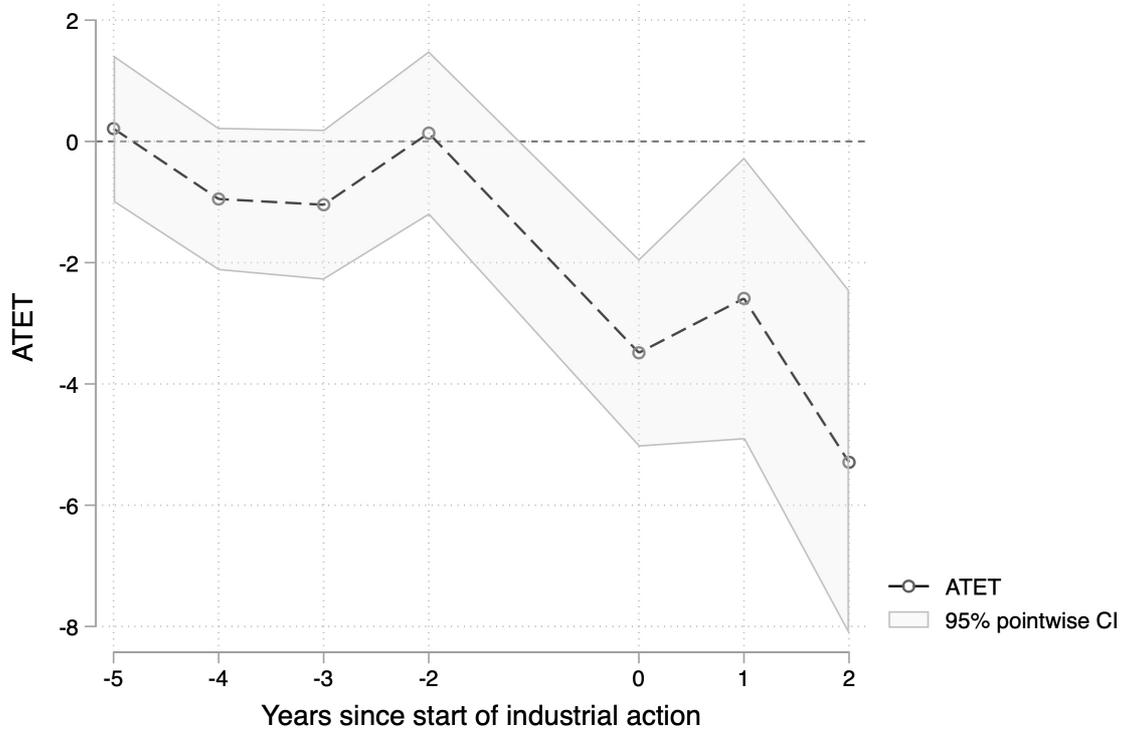

Panel (c) Satisfaction with teaching (NSS)

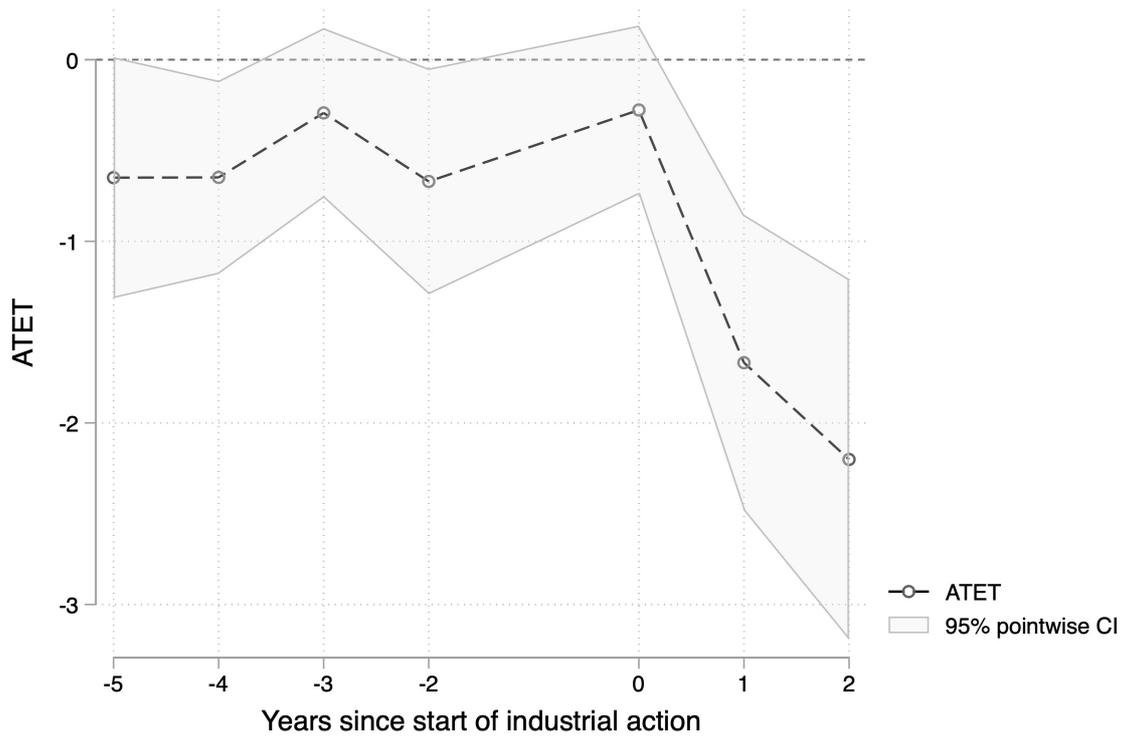

Panel (d) Satisfaction with course (NSS)

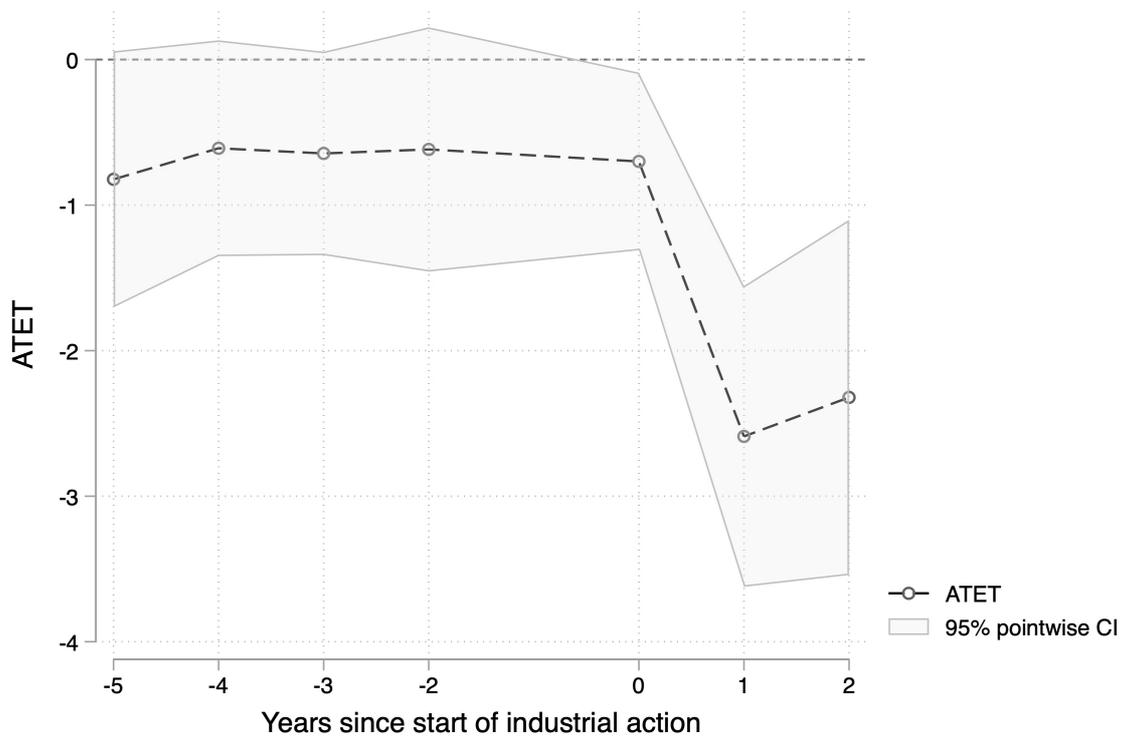

Panel (e) Value added score

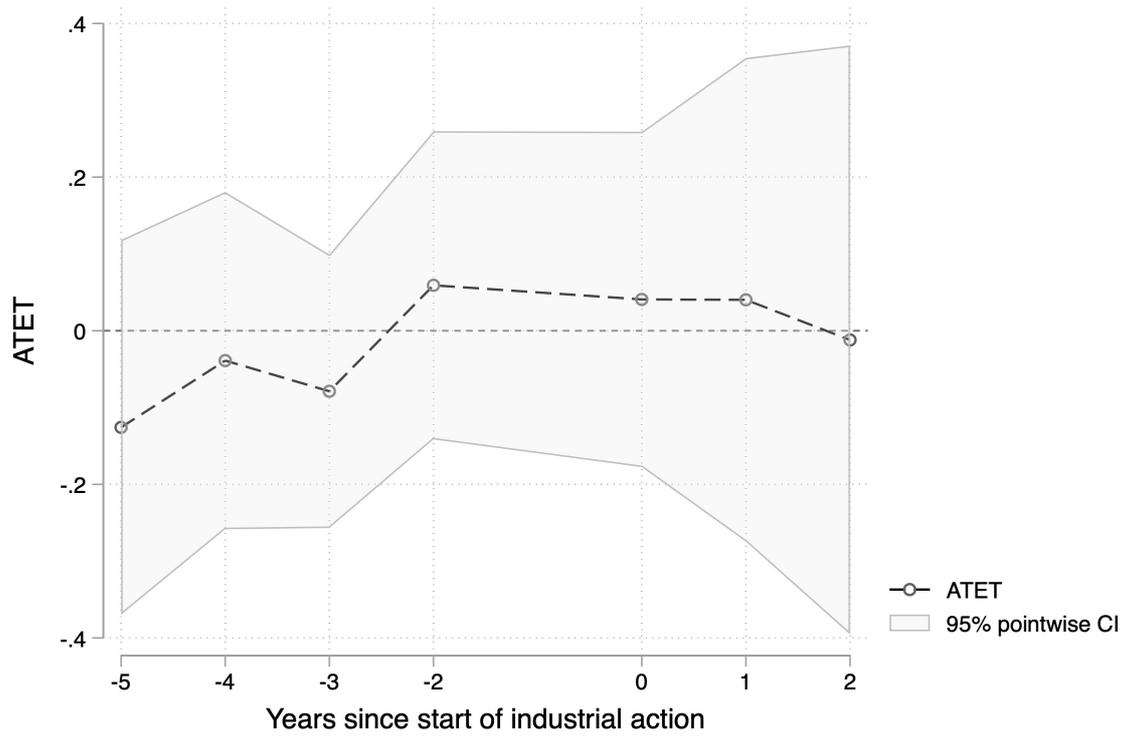

**Figure 5: Heterogeneity by timing of initial industrial action**
Panel (a): Guardian Rank

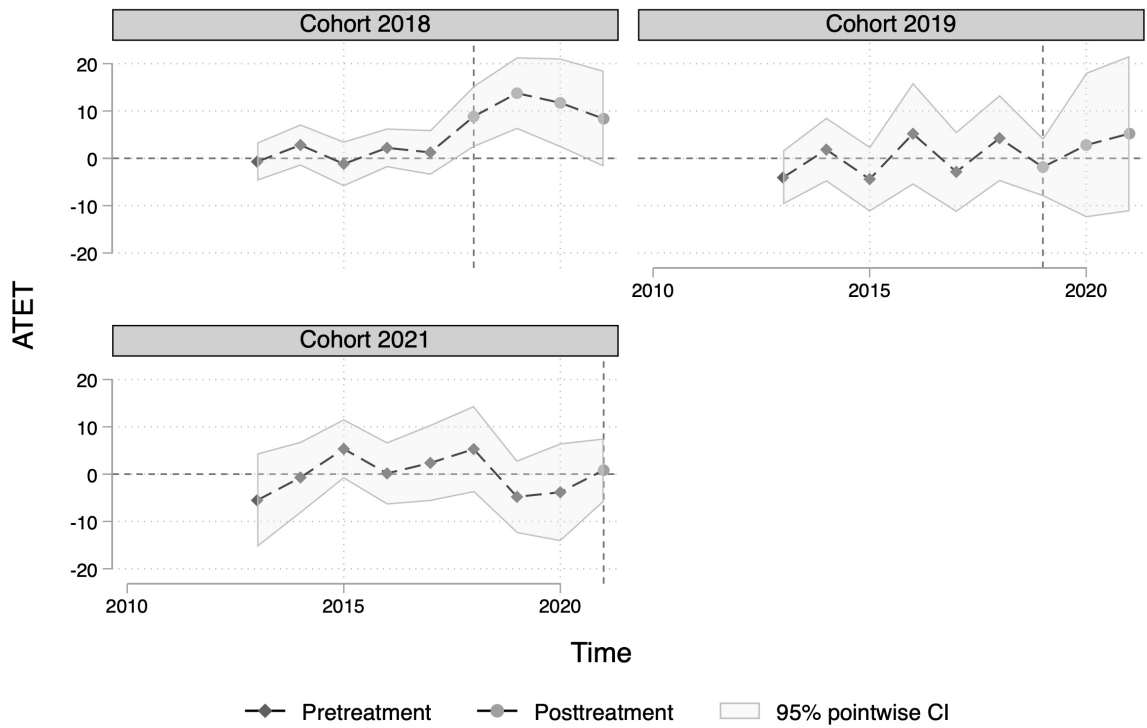

Panel (b) Guardian Score

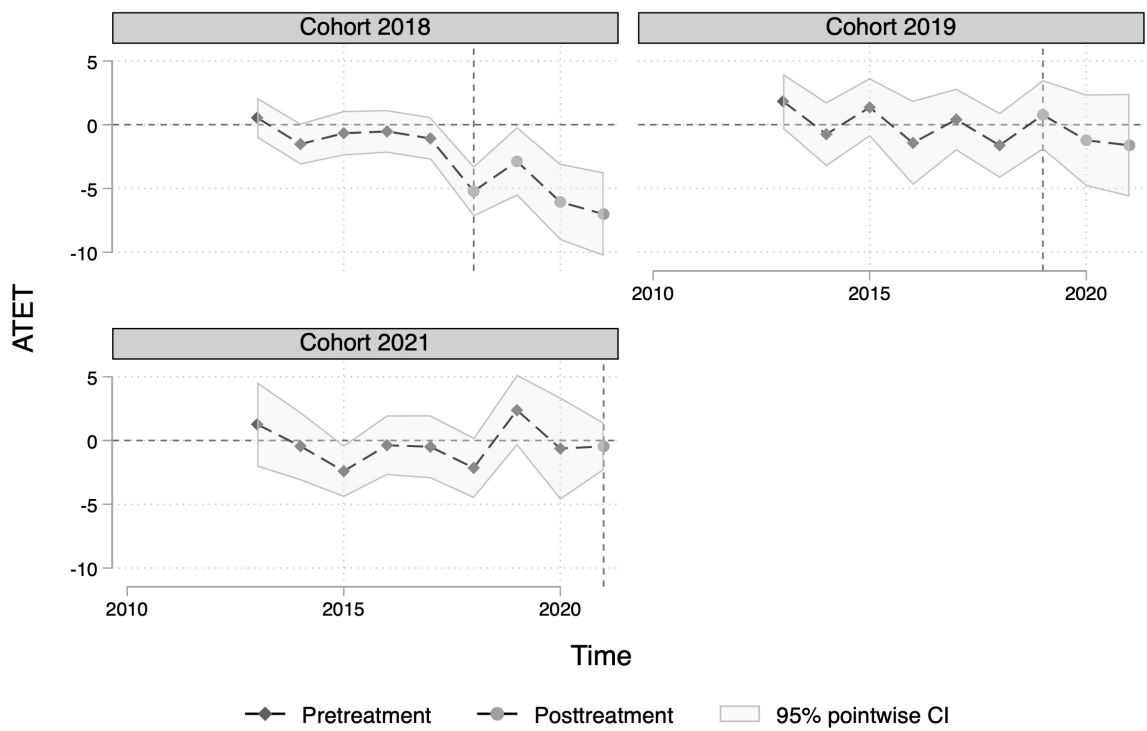

Panel (c) Satisfaction with teaching (NSS)

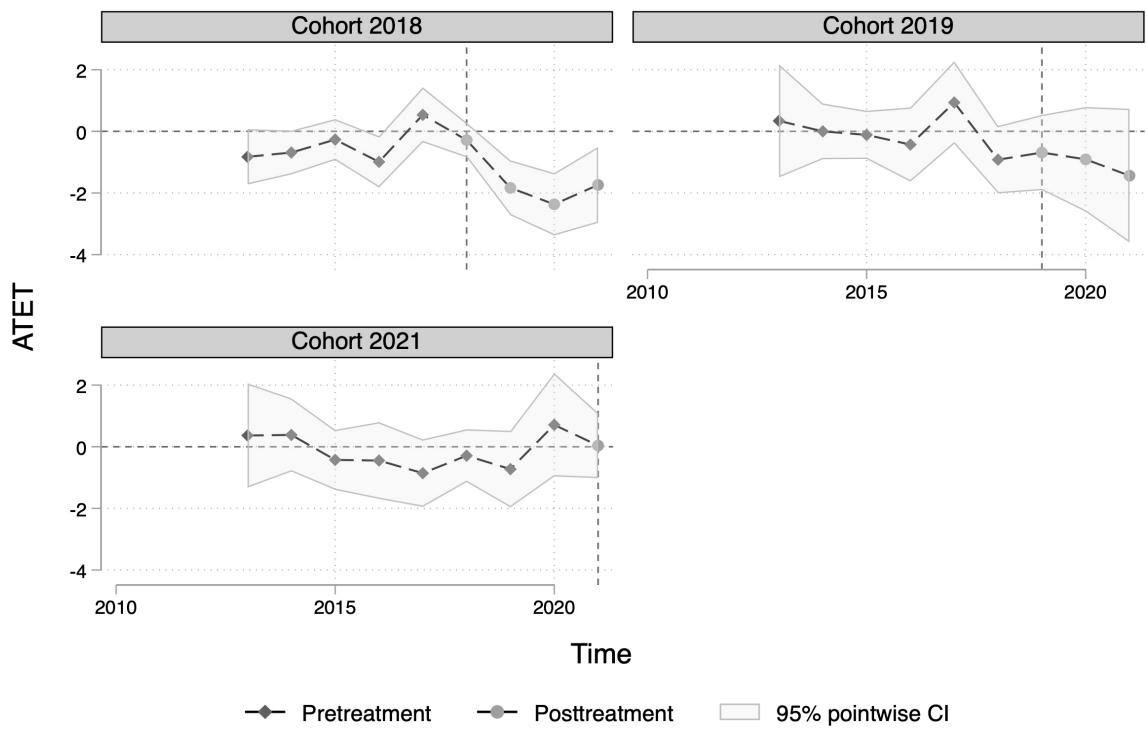

Panel (d) Satisfaction with course (NSS)

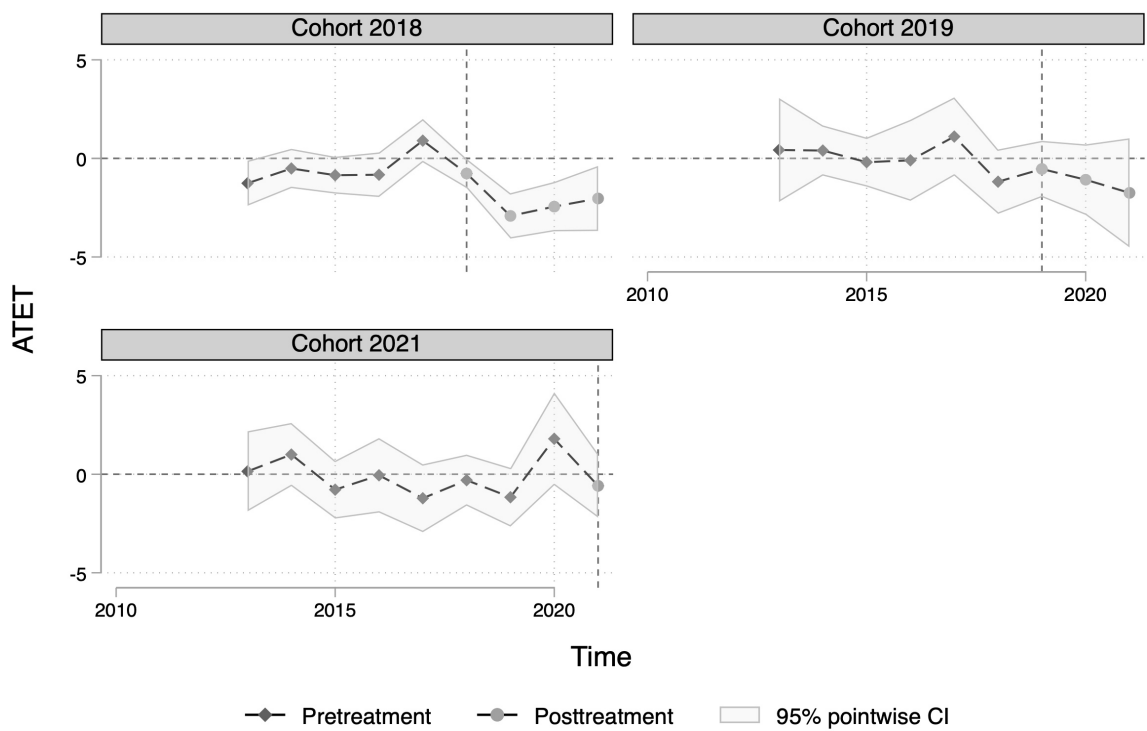

Panel (e) Value added score

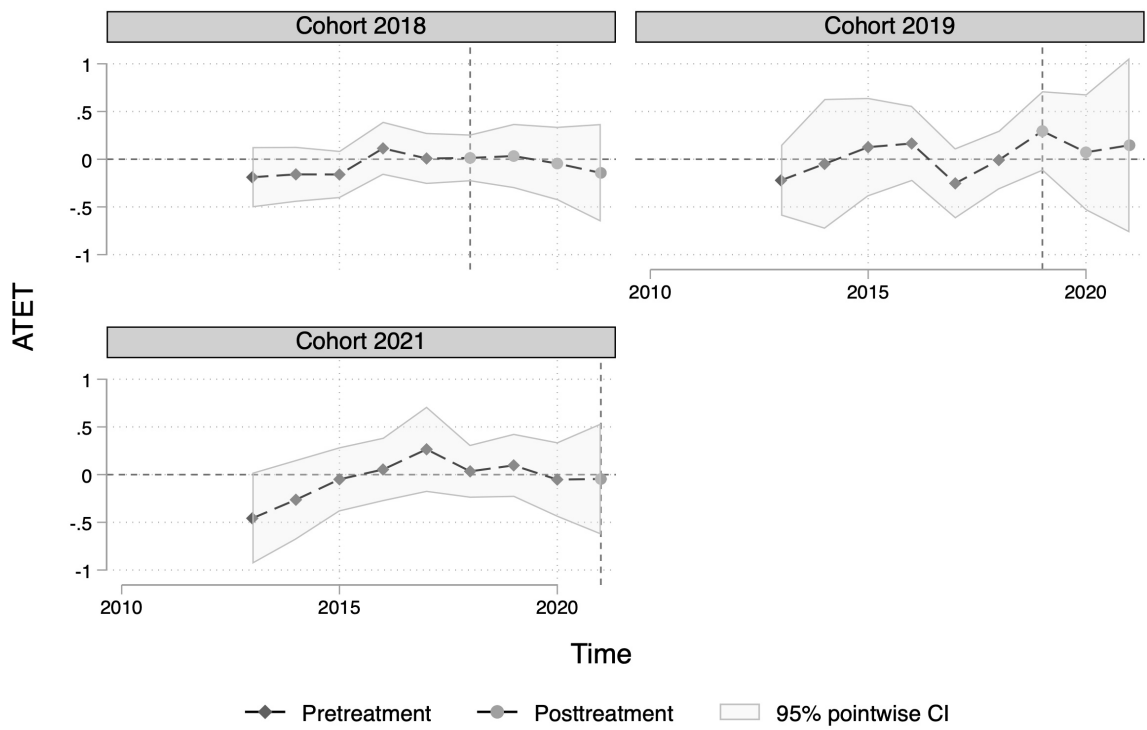